\documentclass[12pt]{article}
																																																																													\pdfoutput=1																																																																											 
\usepackage{amsmath,amssymb}
\usepackage{graphicx}
\usepackage[pdftex]{hyperref}
\DeclareGraphicsExtensions{.eps,.bmp,.wmf,.jpeg,.pdf}
\numberwithin{equation}{section}
\def\be{\begin{equation}}
\def\ee{\end{equation}}

\def\bea{\begin{eqnarray}}
\def\eea{\end{eqnarray}}
\title{$\Lambda$CDM-like models with future singularities}
\author{L.N. Granda\thanks{luis.granda@correounivalle.edu.co} \\ {\small\it Departamento de Fisica, Universidad del Valle}\\{\small\it A.A. 25360, Cali, Colombia}}

\date{}
\begin{document}
\maketitle

\begin{abstract}
\noindent We consider new models of dark energy with finite time future singularities, by introducing the pressure density as a function of the scale factor. This approach gives acceptable phenomenological models of dark energy, practically indistinguishable from the cosmological constant up to the present, which face future singularities at finite time and finite scale factor. Exact scalar field model representation was found for quintessence, Big Rip and type III singularity models. The simple form of the equation of state allows to establish a relationship between its current value, $w_0$, and the time or redshift at which the singularity takes place. The effect on the growth of matter perturbations was calculated.\\ 

\end{abstract}

\maketitle

\section{Introduction}
\noindent 
The most appealing explanation of the accelerated expansion \cite{riess}, \cite{perlmutter}, \cite{kowalski}, \cite{hicken}, \cite{komatsu}, \cite{percival}, \cite{komatsu1}, is considering an exotic fluid with negative pressure, called dark energy (DE), which represents about 70\% of the energy content of the universe and is causing an accelerated expansion that behaves very close to the acceleration given by the cosmological constant. According to astrophysical observations the dark energy equation of state (DE EoS) $w$ lies in a narrow interval around $w=-1$, being consistent with current values slightly bellow this limit. This opens the possibility to an universe that undergoes three phases throughout its evolution. The initial matter-dominance phase with decelerated expansion, followed by the transition to accelerated phase with $-1<w<-1/3$, and the probably current or future transition to the so called phantom phase, characterized by $w<-1$, the weak energy condition is violated.
Several works have been dedicated to the theoretical possibilities of an expanding universe dominated by phantom dark energy. These models \cite{caldwell} lead to a different types of singularities \cite{sergei21, bamba}, the most drastic of which is the so called Big Rip singularity \cite{frampton}, \cite{kamion}, in which the scale factor, the density and pressure become infinite at a finite time. The type II or sudden singularities are characterized by finite $a$ and $\rho$ but divergent $p$ at finite time \cite{barrow}, \cite{barrow1}. In the type III singularities both $\rho$ and $p$ diverge, but the scale factor remains finite at finite time \cite{sergei22}, \cite{stefancic}. Finite-time singularities which are softer than the previous, are classified as type IV singularities in which the scale parameter remains finite and the density and pressure become zero or finite at finite time, but higher derivatives of the Hubble parameter are divergent \cite{sergei21}. 
Models known as Little Rip, in which the DE increases with time as in the Big Rip but without future singularity, have been proposed\cite{frampton1}, \cite{brevik},\cite{granda}.\\
Most of the studied singularities so far, are obtained by introducing explicit dependence of the scale factor on time, or considering phenomenological models with generalized EoS in which the pressure density $p$ of dark energy is given in terms of the energy density as some function $p(\rho)$ \cite{sergei21}, \cite{barrow, barrow1, sergei22, stefancic}, \cite{sergei24, sergei25, barrow2, ruth, granda, astashenok}. Inhomogeneous EoS have been also considered, where the pressure may depend on the Hubble parameter and its derivatives  \cite{nojiri, bamba}, motivated by symmetry considerations or by some generalized gravity theories.
Another interesting way to study the phenomenology of DE could be to define the pressure density through the scale factor. This approach allows to consider singularities that take place at finite scale factor, where this finite value is reached at finite time. The pressure is proposed as $p=f(a)$ and the energy density is found after the integration of the continuity equation. From the continuity equation follows that the state with  EoS $w= -1$ is reached asymptotically as $\dot{\rho}\rightarrow 0$, and the EoS never crosses the phantom divide. Nevertheless when we add a second fluid with constant EoS, which is the homogeneous solution to the continuity equation (in fact is a pressureless fluid $w=0$), then the resulting EoS can cross the phantom divide for certain types of future singularities. 
This allows to introduce a general approach for the construction of dark energy models with different future scenarios, but that from the past and to the present are consistent with the observational data.
In the present work we investigate new DE cosmologies with finite-time singularities, by introducing the pressure density as a function of the scale factor. These models present singularities of the type II-IV characterized by finite scale factor at finite time. The paper is organized as follows.
In section 2 we present the background equations and introduce the DE fluid model describing its relevant properties, and derive the conditions for the future singularities. In section 3 we consider several models with type II-IV singularities with explicit cases describing viable cosmological scenarios. The scalar field model representation of the DE fluid is given under reasonable approximation and one exact case is given. Some summary and conclusions are given in the  discussion section.
\section{The dark energy fluid model}
We will consider a fluid in the spatially flat homogeneous and isotropic FRW background
\be\label{eq1}
ds^2=-dt^2+a(t)^2\sum_{i=1}^3dx_i^2
\ee
the dynamics is determined from the equations
\be\label{eq2}
H^2=\frac{\kappa^2}{3}\rho
\ee
\be
-3H^2-2\dot{H}=\kappa^2 p
\ee
where $\rho$ and $p$ are the energy and pressure densities of the fluid that makes up the universe. From these Eqs. follows the continuity equation
\be\label{eq3}
\dot{\rho}+3H\left(\rho+p\right)=0
\ee
Another way to integrate this equation is by giving the pressure in terms of the scale factor, i.e. $p=p(a)$. As a result of solving the Eq. (\ref{eq3}) we find the density in terms of the scale factor, and therefore determine the equation of state. Having the density in terms of the scale factor we can integrate Eq. (\ref{eq1}), which gives the time dependence of the scale factor and completes the cosmological description of the model . The advantage of this method is that it conduces to realistic cosmologies that include in any solution the early time matter dominance with the energy density behaving as $\rho\propto a^{-3}$. The Eq. (\ref{eq3}) may be written in terms of the scale factor as independent variable, as
\be\label{eq4}
a\frac{d\rho(a)}{da}+3\left[\rho(a)+p(a)\right]=0
\ee
Solving this equation gives $\rho(a)$, which defines the equation of state parameter (EoS) 
\be\label{eq4aa}
w=\frac{p(a)}{\rho(a)}
\ee 
The general solution to this differential equation includes the solution to the homogeneous equation (corresponding to $p=0$), which describes the dust type matter. Then, we assume that the non-homogeneous solution describes the dark energy. The solution to Eq. (\ref{eq4}) together with the pressure $p(a)$, give parametrically the equation of state of the DE fluid (or DE plus matter) through the scale factor as the parameter. Considering the non-homogeneous solution to (\ref{eq4}), we can find the scalar field that represents the DE fluid, from the minimally coupled scalar field model as follows
\be\label{eq3a}
\rho=\pm \frac{1}{2}\dot{\phi}^2+V(\phi)
\ee
\be\label{eq3b}
p=\pm \frac{1}{2}\dot{\phi}^2-V(\phi)
\ee
which give
\be\label{eq3c}
\rho+p=\pm\dot{\phi}^2
\ee
where the minus sign corresponds to the phantom scalar. In therms of the scale factor, this equation can be written as
\be\label{eq3d}
\frac{d\phi}{da}=\pm \frac{1}{aH}\sqrt{|\rho(a)+p(a)|}
\ee
for the scalar field dominated universe, we find (using (\ref{eq2})
\be\label{eq3e}
\phi=\phi_0\pm \frac{\sqrt{3}}{\kappa}\int_{a_0}^a \frac{\sqrt{|\rho(a)+p(a)|}}{a\sqrt{\rho}} da.
\ee
From the Eqs. (\ref{eq3a}) and (\ref{eq3b}) we find the scalar potential as function of the scale factor
\be\label{eq3f}
V=\frac{1}{2}\left[\rho(a)-p(a)\right]
\ee
where the scalar field density and pressure satisfy the continuity equation (\ref{eq4}) which is equivalent to the equation of motion for the scalar field. In this manner, the scalar field can be reconstructed from the given dark energy density and pressure in terms of the scale factor. If the integration in (\ref{eq3e}) can be performed analytically, the scalar field field model is described parametrically in terms of the scale factor, or in some cases it could be possible to find the explicit expression for potential as function of the scalar field.   
In the present work we investigate viable cosmologies with finite time future singularities characterized by finite $a$. As will be shown, in some cases it is possible to find the explicit dependence of the pressure on the energy density $p(\rho)$.\\

\noindent {\bf The Pressure as Function of the Scale Factor}. \\
Let us consider the following expression for the pressure  
\be\label{eq4f}
p=-A\left[1-\left(\frac{a}{a_s}\right)^{\alpha}\right]^{\beta}
\ee
where $A$ and $\alpha$ are positive constants. Replacing in the continuity equation (\ref{eq4}) gives the density
\be\label{eq4g}
\rho=A\hspace{0.1cm} _{2}F_1\left[\frac{3}{\alpha},-\beta,\frac{\alpha+3}{\alpha},\left(\frac{a}{a_s}\right)^{\alpha}\right]
\ee
Replacing $p$ from (\ref{eq4f}) into (\ref{eq4g}) we find the explicit dependence of the density on the pressure:
\be\label{eq4h}
\rho=A \hspace{0.1cm} _{2}F_1\left[\frac{3}{\alpha},-\beta,\frac{\alpha+3}{\alpha},1-\left(-\frac{p}{A}\right)^{1/\beta}\right]
\ee
In some cases, as will be shown, it is possible to invert this equation and express the pressure $p$ in the standard form, as function of the density.
From the above expressions follow the DE EoS as
\be\label{eq4h1}
w_{DE}=-\frac{\left[1-\left(\frac{a}{a_s}\right)^{\alpha}\right]^{\beta}}{\hspace{0.1cm} _{2}F_1\left[\frac{3}{\alpha},-\beta,\frac{\alpha+3}{\alpha},\left(\frac{a}{a_s}\right)^{\alpha}\right]}.
\ee
This expression allow to establish a relationship between the current value of the EoS, $w_0$ and the scale factor at which the singularity takes place. By setting $a=1$, it follows
\be\label{eq4h11}
w_0=-\frac{\left[1-\left(a_s\right)^{-\alpha}\right]^{\beta}}{\hspace{0.1cm} _{2}F_1\left[\frac{3}{\alpha},-\beta,\frac{\alpha+3}{\alpha},\left(a_s\right)^{-\alpha}\right]}=-\frac{\left[1-\left(1+z_s\right)^{\alpha}\right]^{\beta}}{\hspace{0.1cm} _{2}F_1\left[\frac{3}{\alpha},-\beta,\frac{\alpha+3}{\alpha},\left(1+z_s\right)^{\alpha}\right]},
\ee
where we used the redshift relation $a=(1+z)^{-1}$. In some cases this relationship becomes very simple, allowing to express $z_s$ in terms of the observable $w_0$.
As we will see bellow, the general model (\ref{eq4f}) and (\ref{eq4g}) can be reduced to elementary functions in some particular cases, but in its general form allows to analyze the different cosmological scenarios that could take place at the limit $a\rightarrow a_s$, depending on the range of the parameters $\alpha$ and $\beta$. It is worth highlighting that the EoS (\ref{eq4h1}) does not depend on $A$ (as follows from (\ref{eq4f}) and (\ref{eq4g})), and will only depend on $a_s$ once we fix the powers $\alpha$ and $\beta$. An important feature of the model is that for any $\alpha>0$, the pressure and density have the limit
\be\label{eq4h2}
\lim_{a \to 0} {p}=-A,\;\;\;\; \lim_{a \to 0} {\rho}=A,
\ee
where the second limit follows form the properties of the hypergeometric function for $\alpha>0$. This indicates that the DE EoS tends asymptotically to $w_{DE}\rightarrow -1$ at $a\rightarrow 0$, and stays all the time below the phantom divide line without crossing it. This  also follows from the continuity equation 
$$ \dot{\rho}=-3H\left(\rho+p\right)$$ 
from which follows that $\dot{\rho}\rightarrow 0$ implies $p\rightarrow -\rho$, which in the present case is achieved at $a\rightarrow 0$.
Nevertheless the situation is different when we add the matter term $\rho_m$, which is the solution to the homogeneous equation (\ref{eq3}) (i.e. for $p=0$).  In this case the effective (or total) EoS $w$ takes the value $w=-1$ at some point satisfying the condition $\dot{\rho}_m(a_{ph})+\dot{\rho}(a_{pd})=0$, where $p+\rho+\rho_m=0$. Then, the effective EoS crosses the phantom divide from above at $a_{pd}$ (or $t_{pd}$), which is reached before the singularity, i.e. $a_{pd}<a_s$. This case is equivalent to a mixture of two non-interacting perfect fluids, one of which with constant equation of state $w_m=p_m/\rho_m=0$.\\
From (\ref{eq3e}) we find the scalar field as
\be\label{eq1A1}
\phi-\phi_0=\pm \frac{\sqrt{3}}{\kappa}\int_{a_0}^a \frac{\sqrt{|\hspace{0.1cm} _{2}F_1\left[\frac{3}{\alpha},-\beta,\frac{\alpha+3}{\alpha},\left(\frac{a}{a_s}\right)^{\alpha}\right]+\left[1-\left(\frac{a}{a_s}\right)^{\alpha}\right]^{\beta}|}}{a\sqrt{\hspace{0.1cm} _{2}F_1\left[\frac{3}{\alpha},-\beta,\frac{\alpha+3}{\alpha},\left(\frac{a}{a_s}\right)^{\alpha}\right]}} da,
\ee
and the scalar potential as
\be\label{eq1A2}
V=\frac{A}{2}\left[\hspace{0.1cm} _{2}F_1\left[\frac{3}{\alpha},-\beta,\frac{\alpha+3}{\alpha},\left(\frac{a}{a_s}\right)^{\alpha}\right]+\left[1-\left(\frac{a}{a_s}\right)^{\alpha}\right]^{\beta}\right]
\ee

\noindent {\bf Exact Big Rip and quintessence solutions}. \\

\noindent Note that away from the singularity, at $a<<a_s$ (or even at $a<a_s$ if the power $\alpha$ is positive and large enough) one can make the following approximations up to $\left(a/a_s\right)^{\alpha}$ order
\be\label{eq1A3}
p=-A\left[1-\beta\left(\frac{a}{a_s}\right)^{\alpha}\right],\;\;\;\; \rho=A\left[1-\frac{3\beta}{3+\alpha}\left(\frac{a}{a_s}\right)^{\alpha}\right],
\ee
where in the density we used the expansion of the hypergeometric function. A very interesting property of the above power-law expansion is that the obtained pressure and density (\ref{eq1A3}) satisfy the continuity equation, and therefore can be considered as an independent dark energy fluid. 
In this case the scalar field and potential give exact solution and can be found in exact form. Replacing (\ref{eq1A3}) into the integral (\ref{eq3e}) it is found, assuming $\beta<0$
\be\label{eq1A4}
\phi-\phi_0=\frac{2M_p}{\sqrt{\alpha}}\sinh^{-1}\left[\sqrt{-\frac{3\beta}{3+\alpha}}\left(\frac{a}{a_s}\right)^{\alpha/2}\right],
\ee
and the potential 
\be\label{eq1A5}
V=A\left[1+\frac{\alpha+6}{6}\sinh^2\left(\frac{\sqrt{\alpha}}{2M_p}(\phi-\phi_0)\right)\right]
\ee
On the other hand, from (\ref{eq1A3}) follows that for any $\alpha>0$ and $\beta<0$, the Eos is bellow the phantom divide, but there is not singularity at $a_s$ since $p$ and $\rho$ can safely crosse $a_s$. In fact, at $a_s$ the EoS takes the finite value $w_s=(\alpha+3)(\beta-1)/(\alpha-3\beta+3)$, which can be $-1$ only in the trivial cases $\alpha=0$ or $\beta=0$. Since in this case $a_s$ does not play any role, we can set $a_s=1$. The fact that $w<-1$ indicates that there should be some singularity. In order to find out the type of singularity, we need to establish the time dependence of the scale factor. By integrating the Friedmann equation (\ref{eq2}) for the density (\ref{eq1A3}) gives the following scale factor
\be\label{eq1A6}
a(t)=\left(-\frac{\alpha+3}{3\beta}\right)^{1/\alpha}\left[\sinh^2\left(\frac{\alpha\sqrt{A}}{2\sqrt{3}M_p}(t_c-t)\right)\right]^{-1/\alpha}
\ee
An important property of this solution is its invariance under time reflection, $(t-t_c)\rightarrow (t_c-t)$. It can be seen that at $t\rightarrow t_c$ the scale factor $a\rightarrow \infty$ and therefore, from (\ref{eq1A3}), $\rho\rightarrow\infty$, $p\rightarrow -\infty$ and the universe undergoes a Big Rip singularity. Neglecting the matter contribution, the time to the singularity can be evaluated as
\be\label{eq1A7}
t_{BR}-t_0=\frac{H_0^{-1}}{\sqrt{A}}\int_1^{\infty}a^{-1}\left[1-\frac{3\beta}{\alpha+1}a^{\alpha}\right]^{-1/2}da=\frac{2H_0^{-1}}{\alpha\sqrt{A}}\sinh^{-1}\left(\sqrt{-\frac{\alpha+3}{3\beta}}\right)
\ee
where $A$ is measured in units of $3M_p^2H_0^2$. Assuming for instance, $A=1$, $\alpha=3$ and $\beta=-1/2$, one finds $t_{BR}-t_0\sim 0.96 H_0^{-1}$.\\
It can be seen that the potential (\ref{eq1A11}) satisfies the slow-roll conditions for $\Lambda$-like DE. The slow-roll parameter $\epsilon=1/2(V_{,\phi}/V)^2$ takes the value
\be\label{epsi}
\epsilon=\frac{\alpha(\alpha+6)^2\sinh[\frac{\sqrt{\alpha}\phi}{M_p}]^2}{2M_p^2\left(6-\alpha+(\alpha+6)\cosh[\frac{\sqrt{\alpha}\phi}{M_p}]\right)^2}
\ee
and taking into account that the scalar field (\ref{eq1A4}) varies very slowly with $a$, for instance taking $\alpha=3$, then in the interval $a\in (0,1)$ the scalar field $\phi$ varies between $\phi=0$ and $\phi\approx 0.55$, the parameter $\epsilon$ varies between $\epsilon=0$ and $\epsilon\approx 0.56$. In fact, in general, for the entire range of the scalar field $\phi\in (0,\infty)$ the slow roll parameter varies between $0$ and its maximum value $\epsilon=\alpha/2$. \\
Expanding the potential (\ref{eq1A5}) in powers of the scalar field, up to second order one finds the know expression
\be\label{eq1A8}
V\approx A\left[1+\frac{\alpha(\alpha+6)}{24M_p^2}\left(\phi-\phi_0\right)^2\right]
\ee 
which appear in many contexts in particle physics and also have been considered as an example of dark energy fluid with exact solution \cite{dutta, astashenok}. In \cite{astashenok} appears as the scalar field potential for a Little Rip solution.\\
Note that the phantom evolution is not the only possible outcome from the model (\ref{eq1A1}), as follows from its EoS (setting $a_s=1$)
\be\label{eq1A9}
w_{DE}=-\frac{1-\beta a^{\alpha}}{1-\frac{3\beta}{\alpha+3}a^{\alpha}}=-\frac{1-\beta (1+z)^{-\alpha}}{1-\frac{3\beta}{\alpha+3}(1+z)^{-\alpha}}
\ee
If $\alpha<-3$ and $\beta>0$ or $-3<\alpha<0$ and $\beta<0$, then $-1<w_{DE}<-1-\alpha/3$, giving a quintessence model. The scalar field becomes
\be\label{eq1A10}
\phi-\phi_0=\frac{2M_p}{\sqrt{-\alpha}}\arcsin\left[\sqrt{\frac{3\beta}{3+\alpha}}a^{\alpha/2}\right],
\ee
and the potential 
\be\label{eq1A11}
V=A\left[1+\frac{\alpha+6}{6}\sin^2\left(\frac{\sqrt{-\alpha}}{2M_p}(\phi-\phi_0)\right)\right]
\ee
The time dependence of the scale factor is the same given by (\ref{eq1A6}) with $\alpha$ negative, which removes the finite time singularity at $t=t_c$ (in this case we can set $t_c=0$). It is clear also that the integral (\ref{eq1A7}) does not converge for $\alpha<-3$ and $\beta>0$ or $-3<\alpha<0$ and $\beta<0$, indicating the absence of finite time singularity. \\
The slow-roll parameter for the potential (\ref{eq1A11}) is given by the expression
\be\label{epsi1}
\epsilon=-\frac{\alpha(\alpha+6)^2\sin[\frac{\sqrt{-\alpha}\phi}{M_p}]^2}{2M_p^2\left(18+\alpha-(\alpha+6)\cos[\frac{\sqrt{-\alpha}\phi}{M_p}].\right)^2}
\ee
Taking for instance $\alpha=-1/2$, then for $a\in (0,1)$, the scalar field varies between $0$ and $\phi\approx 2.5$ and $\epsilon$ takes its maximum value $\epsilon_{max}\approx 0.027$ at $\phi=2\sqrt{2}\arctan(2\sqrt{3/23})$. Thus, the potential (\ref{eq1A11}) clearly satisfies the slow-roll condition on $\epsilon$. \\
The expression for the EoS (\ref{eq1A9}) allows to compare with the known redshift parametrization of the DE EoS, namely the CPL parametrization \cite{cpl1, cpl2}
\be\label{eq1A12}
w(z)=w_0+w_1\frac{z}{1+z}. 
\ee
Expanding the EoS (\ref{eq1A9}) in powers of $z/(1+z)$ it is found
\be\label{eq1A13}
w_{DE}=\frac{(\alpha+3)(\beta-1)}{\alpha-3\beta+3}-\frac{\alpha^2(\alpha+3)\beta}{\left(\alpha-3\beta+3\right)^2}\frac{z}{1+z}+...
\ee
comparing both expressions allow to write $\alpha$ and $\beta$ in terms of the CPL parameters as
\be\label{eq1A14}
\alpha=-\frac{3w_0^2+6w_0+w_1+3}{w_0+1},\;\;\;\; \beta=\frac{(w_0+1)(3w_0^2+3w_0+1)}{w_1}
\ee
that can be used as a criteria of consistency of the model with observations at high redshift and are also useful for linear approximation at low redshift. The above expansion is valid only when there is some fast convergence criterion. Analyzing higher order terms in the expansion (\ref{eq1A13}), we find that this criterion is fulfilled in the cases when $0<\alpha\le 3$ and $-1/2\le\beta<0$ (phantom) or $-1<\alpha<0$ and $-1/2<\beta<0$ (quintessence). Only in that cases the CPL parametrization is useful for the models (\ref{eq1A3}). There are also important criteria, as discussed below, useful to evaluate the departure of the model from the cosmological constant.
\section{Finite $a$ Singularities}
In this section we consider different types of future singularities with finite scale factor \cite{barrow, stefancic, nojiri}, assuming values for $\alpha$ and $\beta$ in the appropriate intervals. In all models the DE EoS maintains very close to the cosmological constant and the total EoS gives acceptable description of the evolution from the early matter dominance to the current time characterized by predominance of the dark energy, being quite  consistent with current observations.
Taking the limit at $a\rightarrow a_s$ in  (\ref{eq4f}) and  (\ref{eq4g}) one can distinguish the following singularities
\begin{enumerate}
\item Type II (sudden) singularities.\\
For any $\alpha>0$ and $-1<\beta<0$ one finds the limits
\be\label{eq4i}
\lim_{a \to a_s} {p}=-\infty,\,\,\,\,\,\,  \lim_{a \to a_s} {\rho}=A\frac{\Gamma(1+\frac{3}{\alpha})\Gamma(\beta+1)}{\Gamma(1+\frac{3}{\alpha}+\beta)},\,\,\, \lim_{a \to a_s} {w}=-\infty
\ee
\item Type III singularities.\\
This singularity takes place for $\beta\leq -1$ and any $\alpha>0$
\be\label{eq4j}
\lim_{a \to a_s} {p}=-\infty ,\,\,\,\,\,\,\, \lim_{a \to a_s} {\rho}=\infty,\,\,\,\, \lim_{a \to a_s} {w}=-\infty
\ee
\item Type IV singularities.\\
Are obtained for any $\alpha>0$ and $0<\beta<1$, giving the limits
\be\label{eq4k}
\begin{aligned}
\lim_{a \to a_s} {p}=0,\,\,\,\, \lim_{a \to a_s} {\rho}=&A\frac{\Gamma(1+\frac{3}{\alpha})\Gamma(\beta+1)}{\Gamma(1+\frac{3}{\alpha}+\beta)},\,\,\, \lim_{a \to a_s} {\left|\frac{d^n p}{da^n}\right|}=\infty, \,\, n=1,2,..\\
& \lim_{a \to a_s} {w}=0
\end{aligned}
\ee
\end{enumerate}
Including the homogeneous solution of (\ref{eq4}), which corresponds to dust type matter, the total density (that includes both, DE and dark matter fluids), takes the form
\be\label{eq4m}
\rho=\frac{\rho_{m0}}{a^3}+A\hspace{0.1cm} _{2}F_1\left[\frac{3}{\alpha},-\beta,\frac{\alpha+3}{\alpha},\left(\frac{a}{a_s}\right)^{\alpha}\right]
\ee  
where $\rho_{m0}$ is the integration constant, associated with the density of dark matter. Replacing $a$ from (\ref{eq4f}) into the above expression for $\rho$, it is found $\rho(p)$ as
\be\label{eq4n}
\rho=\frac{\rho_{m0}}{a_s^3}\left[1-\left(-\frac{p}{A}\right)^{1/\beta}\right]^{-3/\alpha}+A\hspace{0.1cm} _{2}F_1\left[\frac{3}{\alpha},-\beta,\frac{\alpha+3}{\alpha},1-\left(-\frac{p}{A}\right)^{1/\beta}\right]
\ee  
Note that the addition of the matter term does not change the character of the singularities.  Replacing the density (\ref{eq4m}) into the Friedmann equation (\ref{eq2}) we find the time remaining to the singularities as
\be\label{eq4l}
t_s-t_0=H_0^{-1}\int_1^{a_s}a^{-1}\left[\frac{\rho_{m0}}{a^3}+A\hspace{0.1cm} _{2}F_1\left[\frac{3}{\alpha},-\beta,\frac{\alpha+3}{\alpha},\left(\frac{a}{a_s}\right)^{\alpha}\right]\right]^{-1/2}da
\ee
where $\rho_{m0}$ and $A$ are measured in units of $3M_p^2H_0^2$ ($H_0$ is the current value of the Hubble parameter). 
Note that the last limit in Eq. (\ref{eq4k}) leads to 
\be
\lim_{t \to t_s} {\left|\frac{d^n H}{dt^n}\right|}=\infty, n=2,3,.., 
\ee
Bellow we consider explicit cases of these singularities, for specific values of $\alpha$ and $\beta$ in the intervals considered above.
\subsection*{Sudden Singularities}
The sudden singularity is a finite time singularity that occurs when the scale factor $a(t)$, its time derivative and the energy density remain finite, while the pressure $p\rightarrow -\infty$, and therefore, the EoS also suffers the singularity, $w\rightarrow -\infty$. To illustrate this type of singularities we consider the following two cases of the model (\ref{eq4f}). \\

\noindent \textbf{ 1.} $\alpha=3$, $\beta=-1/2$\\
The model (\ref{eq4f}) gives
\be\label{eq6}
p=-A\left[1-\left(\frac{a}{a_s}\right)^3\right]^{-1/2},\,\,\,\,\, \rho=2A\left(\frac{a_s}{a}\right)^3\left(1-\sqrt{1-\left(\frac{a}{a_s}\right)^3}\right)
\ee
combining these two equations one finds 
\be\label{eq7}
p=\frac{A\rho}{\rho-2A}
\ee
And the equation of state takes the simple form
\be\label{eq7a}
w_{DE}=\frac{A}{\rho-2A}=\frac{1}{2\left[\left(\frac{a_s}{a}\right)^3\left(1-\sqrt{1-\left(\frac{a}{a_s}\right)^3}\right)-1\right]}
\ee
The sudden singularity takes place when $\rho\rightarrow 2A$ as follows from (\ref{eq6}) (at $a\rightarrow a_s$) or (\ref{eq7}) at $\rho\rightarrow 2A$. Note that away form the singularity, when $a<<a_s$, one can approximate $\rho$ as
\be\label{eq7b}
\rho\simeq 2A\left(\frac{a_s}{a}\right)^3\left[1-\left(1-\frac{1}{2}\left(\frac{a}{a_s}\right)^3\right)\right]=A
\ee
Then the fluid (\ref{eq6}), under the condition $a<<a_s$, behaves very similar to the cosmological constant with $p\simeq -A(1+\frac{1}{2}(a/a_s)^3)\sim -A$ and $\rho\simeq A$ ($w\sim -1$), and the further the singularity is, the closer the behavior to the cosmological constant. In fact, taking the limit $a_s\rightarrow \infty$ in (\ref{eq7a}) one finds $w=-1$. \\
By adding the homogeneous solution to the density in (\ref{eq6}), we find the total density as
\be\label{eq7e}
 \rho_t=\frac{\rho_{m0}}{a^3}+2A\left(\frac{a_s}{a}\right)^3\left(1-\sqrt{1-\left(\frac{a}{a_s}\right)^3}\right)
 \ee
combining this result with the equation (\ref{eq6}) for the pressure, one finds
\be\label{eq7f}
p=-\frac{A^2}{\rho_c-\rho_t}\left(1-\sqrt{1-\frac{\rho_t}{A^2}\left(\rho_c-\rho_t\right)}\right)
\ee
where $\rho_c=2A+\frac{\rho_{m0}}{a_s^3}$. The sudden singularity takes place when $\rho\rightarrow \rho_c$ (at $a\rightarrow a_s$) as follows from (\ref{eq7e}) or (\ref{eq7f}). \\
\noindent \textbf{2.} $\alpha=6$, $\beta=-1/2$\\
From (\ref{eq4f}) it follows that
\be\label{eq8}
p=-A\left[1-\left(\frac{a}{a_s}\right)^6\right]^{-1/2},\,\, \rho=A\left(\frac{a_s}{a}\right)^3 \arcsin\left[\left(\frac{a}{a_s}\right)^3\right]
\ee
combining these equations gives $\rho$ in terms of $p$ as
\be\label{eq9}
\rho=\frac{1}{\sqrt{1-\left(\frac{A}{p}\right)^2}}\left(A\arcsin\left[\sqrt{1-\left(\frac{A}{p}\right)^2}\right]\right)
\ee
Taking the limit $a\rightarrow a_s$, we find $p\rightarrow -\infty$, $\rho\rightarrow A\pi/2$ and $w\rightarrow-\infty$, corresponding to sudden singularity. 
If $p/A\simeq -1$  then from (\ref{eq9}) follows that $\rho\simeq A$ and $w\simeq -1$, and away from the singularity the model is very close to the cosmological constant. 
By adding solution of the homogeneous equation (\ref{eq4}) to the density (\ref{eq8}), the total density becomes
\be\label{eq9d}
 \rho_t=\frac{\rho_{m0}}{a^3}+A\left(\frac{a_s}{a}\right)^3 \arcsin\left[\left(\frac{a}{a_s}\right)^3\right],
 \ee
or combining with the equation for $p$ in (\ref{eq8})
\be\label{eq9e}
\rho_t=\frac{1}{\sqrt{1-\left(\frac{A}{p}\right)^2}}\left(\frac{\rho_{m0}}{a_s^3}+A\arcsin\left[\sqrt{1-\left(\frac{A}{p}\right)^2}\right]\right)
\ee
From the Friedmann equation for the densities (\ref{eq7e}) and (\ref{eq9e}), and applying the flatness condition (at $a=1$), writing $\rho_{m0}$ and $A$ in units of ($3M_p^2H_0^2$), we can determine the constant $A$, provided $\rho_{m0}=0.3$ and given $a_s$. Then we can compose the equation of state for each case in terms of the redshift using $a=(1+z)^{-1}$. 
Note that due to the high power of $a/a_s$ for the above two cases, it is not necessary to maintain the condition $a_s>>1$, and it is enough with $a_s>1$. Taking for instance $a_s=50$ (assumed for $a_s>>1$) and $a_s=10$ or even $a_s=5$ (that satisfy $a_s>1$), give the effective EoS practically indistinguishable from the $\Lambda$CDM as shown in Fig. 1. Applying the flatness condition, at $a=1$, to the densities (\ref{eq7e}) and (\ref{eq9d}) gives $A\approx 0.7$ for all the cases. 
In Fig. 1 we show the evolution of the effective EoS for the cosmological scenarios with $a_s=50$, $a_s=10$ and $a_s=5$, for the above models leading to sudden singularity.
\begin{center}
\includegraphics [scale=0.7]{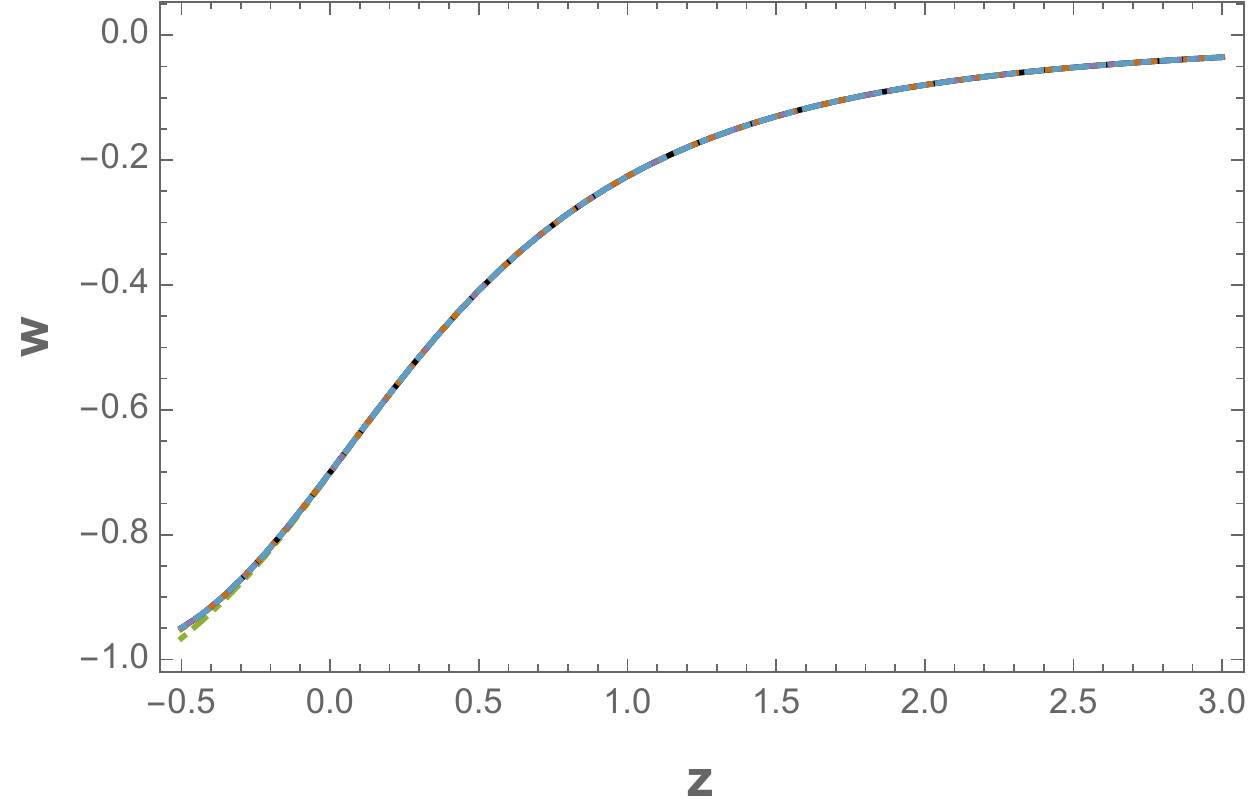}
\end{center}
\begin{center}
{Fig. 1 \it The behavior of the EoS for the $\Lambda$CDM and for the two cases of sudden singularities, for the cosmological scenarios with $a_s=50$, $a_s=10$ and $a_s=5$. The models correspond to $(\alpha=3,\beta=-1/2)$ and  $(\alpha=6,\beta=-1/2)$. In all cases the EoS are practically indistinguishable from each other and from the $\Lambda$CDM}.
\end{center}
All the curves are practically indistinguishable between them and from the behavior of $\Lambda$CDM up to the present (and even in the future, up to some time before the singularity), except that they face a sudden singularity at different times at the future. How far is this future, depends on the value of the scale factor at which the singularity takes place, $a_s$. The time to the singularity for the considered cases is shown in Fig. 1. Here we used the flatness condition at $a=1$  with $A$ and $\rho_{m0}$ measured in units of $3H_0^2M_p^2$, and $\rho_{m0}=0.3$
\begin{center}
\begin{tabular}{|p{2cm}|p{2cm}|p{2cm}|}\hline
\multicolumn{3}{|c|}{ Time to the sudden singularity in units of $H_0^{-1}$}\\ \hline
$a_s$ & (3,-1/2) &(6,-1/2)\\ \hline
50  & 4.54& 4.58 \\ \hline
10& 2.61 & 2.65  \\ \hline
5& 1.79 & 1.85 \\ \hline
\end{tabular}
\end{center}
\begin{center}
{Table. 2 \it The time remaining to the type III singularity. The numbers in parenthesis correspond to the values ($\alpha,\beta$) for each model.} 
\end{center}  
\subsection*{Type III Singularities}
The following cases lead to the limit at  $a\rightarrow a_s$: $p\rightarrow -\infty$, $\rho\rightarrow \infty$ and $w\rightarrow -\infty$. \\

\noindent \textbf{1.} $\alpha=3,\,\, \beta=-1$\\
\be\label{eq11}
p=-\frac{A}{1-\frac{a^3}{a_s^3}},\,\,\, \rho=-A\frac{a_s^3}{a^3}\ln\left(1-\frac{a^3}{a_s^3}\right)
\ee
eliminating $a$ we find
\be\label{eq12}
\rho=-A\left(1+\frac{A}{p}\right)^{-1}\ln\left(-\frac{A}{p}\right)
\ee
Proceeding as in the previous cases, and adding the matter term to the density (\ref{eq11}) it is found for the total density
\be\label{eq12c}
\rho_t=\frac{\rho_{m0}}{a^2}-A\frac{a_s^3}{a^3}\ln\left(1-\frac{a^3}{a_s^3}\right)
\ee
or in terms of the pressure
\be\label{eq12e}
\rho_t=\left(1+\frac{A}{p}\right)^{-1}\left[\frac{\rho_{m0}}{a_s^3}-A\ln\left(-\frac{A}{p}\right)\right].
\ee

\noindent \textbf{2.} $\alpha=3,\,\,\, \beta=-2$\\
\be\label{eq13}
p=-\frac{A}{\left(1-\frac{a^3}{a_s^3}\right)^2},\,\,\, \rho=\frac{A}{1-\frac{a^3}{a_s^3}}
\ee
eliminating $a$ gives the pressure explicitly in terms of the density as  
\be\label{eq14}
p=-\frac{\rho^2}{A}. 
\ee
And the DE equation of state is given by
\be\label{eq14a}
w_{DE}=-\frac{\rho}{A}=-\frac{1}{1-\frac{a^3}{a_s^3}}
\ee
From (\ref{eq13}) follows that $\rho/A>1$, and therefore $w<-1$. The EoS depends only on $a_s$ an can be kept in a very narrow region below the cosmological constant, as will be illustrated below. This simple form of the EoS gives a relationship between the current EoS and the singular value $a_s$. If $w_0$ is the value of $w_{DE}$ at $a=1$, then
\be\label{eq14A1}
a_s=\left(\frac{w_0}{w_0+1}\right)^{1/3}
\ee
thus for instance, if $w_0\approx -1.05$, then $a_s\approx 2.76$. The time to the singularity, evaluating the integral (\ref{eq4l}) for this case with $a_s=2.76$ and setting $\rho_{m0}=0$, gives $t_s-t_0\approx 0.84 H_0^{-1}$ (the contribution of matter increases a little this value).\\
The DE fluid (\ref{eq13}) allows the integration in (\ref{eq3e}) and we can find the exact form of the scalar field and the potential that produce the EoS (\ref{eq14a}). After integration in (\ref{eq3e}) with de density and pressure given by (\ref{eq13}) we find the scalar field
\be\label{eq14b}
\phi=\frac{2}{\sqrt{3}}M_p\arcsin\left(\frac{a}{a_s}\right)^{3/2},
\ee
and the potential 
\be\label{eq14c}
V=\frac{A}{2}\frac{\cos^2\left(\frac{\sqrt{3}}{2}\frac{\phi}{M_p}\right)+1}{\cos^4\left(\frac{\sqrt{3}}{2}\frac{\phi}{M_p}\right)}
\ee
where the maximum value of $\phi$ is $\phi_s=\pi M_p/\sqrt{3}$, and at $\phi_s$ the type III singularity takes place. As in the previous cases, by adding the matter content to the density in (\ref{eq13}) one can write the total density as
\be\label{eq14d}
\rho_t=\frac{\rho_{m0}}{a^3}+\frac{A}{1-\frac{a^3}{a_s^3}}.
\ee
By eliminating $a$ between the pressure $p$ given in (\ref{eq13}) and the density $\rho$ in (\ref{eq14d}) we can express the pressure as function of the density as follows
 \be\label{eq14e} 
p=-\frac{1}{2A}\left[\rho_c^2+\rho_t^2-2\frac{\rho_{m0}}{a_s^3}\rho_t+\sqrt{\left(\rho_c^2+\rho_t^2-2\frac{\rho_{m0}}{a_s^3}\rho_t\right)^2-4A^2\rho_t^2}\right]
\ee
where $\rho_c=A-\frac{\rho_{m0}}{a_s^3}$. It can be checked in this expression that if $A=0$, then $\rho_t=\rho_{m0}/a^3$ and $p=0$. \\
The slow-roll parameter for the potential (\ref{eq14c}) is given by the expression
\be\label{epsi3}
\epsilon=\frac{3\left(\cos[\sqrt{3}\phi]+5\right)^2\tan[\sqrt{3}\phi/2]^2}{2\left(\cos[\sqrt{3}\phi]+3\right)^2}.
\ee
Assuming $as=5$, then for $a\in (0,1)$, the scalar field varies between $\phi=0$ and $\phi=\frac{2}{\sqrt{3}}\arcsin[\frac{1}{5\sqrt{5}}]\approx 0.1$, and the slow-roll parameter varies between $\epsilon=0$ and $\epsilon\approx 0.025$. This small value of $\epsilon$ guarantees the quasi-$\Lambda$ behavior of the DE. 

\noindent \textbf{3.} $\alpha=6,\,\,\, \beta=-1$\\
\be\label{eq15}
p=-A\left[1-\frac{a^6}{a_s^6}\right]^{-1},\,\,\, \rho=A\frac{a_s^3}{a^3} \tanh^{-1}\left[\frac{a^3}{a_s^3}\right]
\ee
or
\be\label{eq16}
\rho=\frac{A}{\sqrt{1+\frac{A}{p}}}\left(\tanh^{-1}\left[\sqrt{1+\frac{A}{p}}\right]\right).
\ee
The total density is obtained by adding the homogeneous solution to the density (\ref{eq15}), giving
\be\label{eq16c}
\rho_t=\frac{\rho_{m0}}{a^3}+A\frac{a_s^3}{a^3} \tanh^{-1}\left[\frac{a^3}{a_s^3}\right]=\frac{1}{\sqrt{1+\frac{A}{p}}}\left(\frac{\rho_{m0}}{a_s^3}+A\tanh^{-1}\left[\sqrt{1+\frac{A}{p}}\right]\right),
\ee
which allows to follow the evolution of the effective EoS. \\
\noindent \textbf{4.} $\alpha=1,\,\,\, \beta=-4$\\
\be\label{eqNA1}
p=-\frac{A}{\left(1-\frac{a}{a_s}\right)^4},\;\;\; \rho=\frac{A}{\left(1-\frac{a}{a_s}\right)^3},
\ee
giving the relationship 
\be\label{eqNA2}
p=-\frac{1}{A^{1/3}}\rho^{4/3}.
\ee
Here the equation of state takes the simple form
\be\label{eqNA3}
w_{DE}=-\left(\frac{\rho}{A}\right)^{1/3}=-\frac{1}{1-\frac{a}{a_s}}=-\frac{1}{1-\frac{1+z_s}{1+z}}.
\ee
The integral (\ref{eq1A1}) is exact in this case and gives the scalar field as
\be\label{eqNA4}
\phi-\phi_0=2\sqrt{3}M_p\arctan \left[\frac{\sqrt{a}}{\sqrt{a_s-a}}\right]
\ee
which gives for the scale factor
$$ a=a_s\sin^2\left(\frac{\phi}{2\sqrt{3}M_p}\right),$$
leading to the potential
\be\label{eqNA5}
V=\frac{A}{2}\left[\left(\sec \phi\right)^6+\left(\sec\phi\right)^8\right].
\ee
The EoS (\ref{eqNA3}) gives a simple relationship between its current value and the critical value of the scale factor. If $w_{DE}(z=0)=w_0$, then
\be\label{eqNA6}
w_0=\frac{1}{z_s}
\ee
Thus, if the DE EoS has the current value $w_0\approx -1.05$, then $z_s\approx -0.95$ ($a_s\approx 20$). The integral (\ref{eq4l}) can be evaluated in exact form in the DE dominant case (setting $\rho_{mo}=0$) giving the following expression for the remaining time to the singularity
\be\label{eqNA7}
t_s-t_0=\frac{6a_c^{3/2}\arctan\left[\sqrt{\frac{a_s-1}{a_s}}\right]+2(1-4a_s)\sqrt{a_s-1}}{3\sqrt{A}\left(a_s\right)^{3/2}}H_0^{-1}.
\ee
Assuming ($a_s=20$), it is found $t_s-t_0\approx 1.93 H_0^{-1}$ ($A$ is evaluated using the flatness condition, giving $A\approx 0.86$). Unlike the previous cases, the EoS (\ref{eqNA3}) may differ appreciably from the cosmological constant at low redshift. In Fig 2 we show some cases
\begin{center}
\includegraphics [scale=0.7]{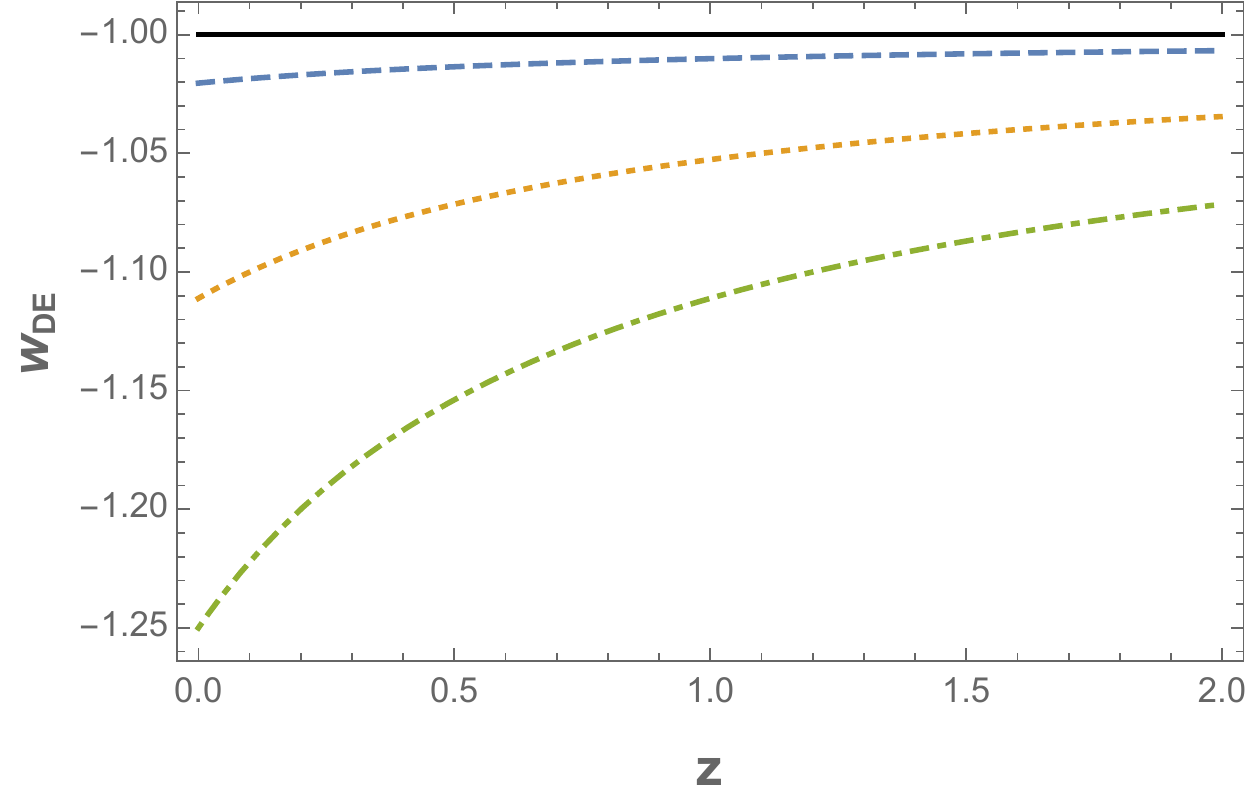}
\end{center}
\begin{center}
{Fig. 2 \it The DE EoS for the model (\ref{eqNA1}) for three scenarios that differ in the scale factor at which the singularity takes place, compared to the cosmological constant (solid line), $a_s=50$ (dashed), $a_s=10$ (dotted), $a_s=5$ (dotdashed).} 
\end{center}
The slow-roll parameter for the potential (\ref{eqNA5}) has the following form
\be\label{epsi4}
\epsilon=\frac{2\left(4\sec[\phi]^2+3\right)^2\tan[\phi]^2}{\left(\sec[\phi]^2+1\right)^2}.
\ee
In this case, the evolution of the scalar field depends strongly on the singular value of the scale factor $a_s$, and the closer the singularity is (smaller $a_s$), the faster the scalar field evolves. Thus, if one assumes $a_s=10$, then in the interval $a\in (0,1)$ the scalar field takes values in the interval $(0, 1.1)$ and the slow-roll parameter varies between $\epsilon=0$ and $\epsilon\approx 122$, and assuming for instance $a_s=10^3$, then in the interval  $a\in (0,1)$  the scalar field evolves in the interval $(0,0.11)$ and $\epsilon$ varies between $0$ and $0.3$. This indicates that the rapid evolution of the scalar field causes a rapid departure from the cosmological constant. Note however, that, even for small $a_s$ according to Fig. 2, the departure from the cosmological constant is slow and not as rapid as the slow-roll parameter suggests. \\
\noindent By adding the mater content, the total density becomes
\be\label{eqNA8}
\rho_t=\frac{\rho_{m0}}{a^3}+\frac{A}{\left(1-\frac{a}{a_s}\right)^3}=\frac{\rho_{m0}}{a_s^3\left(1-\left(-\frac{A}{p}\right)^{1/4}\right)^3}+A\left(-\frac{p}{A}\right)^{3/4}
\ee
The analysis shows that the behavior of the effective EoS for the first three cases is exactly the same as the obtained for the two cases of the sudden singularity. So, Fig. 1 also describes the EoS for the type III singularity models (\ref{eq11}), (\ref{eq13}) and (\ref{eq15}) (at least up to the present), and apart from the character of the singularity, they differ in the time at which the singularity takes place, which are also similar. The special case is the model (\ref{eqNA1}) which has an appreciable difference (in the range of current observations) at low redshift with the cosmological constant. This model has exact representation in terms of the phantom scalar field with potential given by (\ref{eqNA5}). In table 2 we show the time to the singularity for the considered cases
\begin{center}
\begin{tabular}{|p{2cm}|p{2cm}|p{2cm}|p{2cm}|p{2cm}|}\hline
\multicolumn{5}{|c|}{Time to the type III singularity in units of $H_0^{-1}$}\\ \hline
$a_s$ & (3,-1) &(3,-2)& (6,-1)&(1,-4)\\ \hline
50  & 4.47& 4.35 & 4.55 & 3.21\\ \hline
10& 2.55 & 2.43 & 2.63 & 1.58\\ \hline
5& 1.72 & 1.61& 1.8 & 0.99\\ \hline
\end{tabular}
\end{center}
\begin{center}
{Table. 2 \it The time remaining to the type III singularity. The numbers in parenthesis correspond to the values ($\alpha,\beta$) for each model.} 
\end{center}

\noindent \textbf{The $q$ and $j$ parameters}.\\

\noindent From the above results, reflected in Fig. 1 for sudden and type III singularities, it follows that it is not possible to distinguish between the above models and the $\Lambda$CDM.\\
As a possible way to break this degeneracy, we can consider other cosmological parameters like the deceleration parameter $q$, or quantities with higher derivatives of the Hubble parameter like the jerk parameter $j$, that could be more sensitive to the model parameters and thus reveal differences with $\Lambda$CDM. The deceleration and jerk parameters are defined as
\be\label{eq16d}
q=-\frac{1}{aH^2}\frac{d^2 a}{dt^2}=\frac{1}{H^2}\left[\frac{1}{2}(1+z)\frac{dH^2}{dz}-H^2\right],
\ee
\be\label{eq16e}
j=\frac{1}{aH^3}\frac{d^3 a}{dt^3}=\frac{1}{2H^2}\left[(1+z)^2\frac{d^2 H^2}{dz^2}-2(1+z)\frac{dH^2}{dz}+2H^2\right].
\ee
The current values of these parameters for the $\Lambda$CDM model are
\be\label{eq16f}
q_0=\frac{3}{2}\rho_{m0}-1,\;\;\; j_0=1
\ee
where $\rho_{m0}$ is measured in units of $3M_p^2 H_0^2$. In fact $j=1$ for the $\Lambda$CDM model all the time since the matter dominated epoch and at the future. These will be the reference values to compare with the models  (\ref{eq7e}), (\ref{eq9d}), (\ref{eq12c}), (\ref{eq14d}) and (\ref{eq16c}). Specially the simplicity of $j$ for the $\Lambda$CDM allows to measure the deviation from $\Lambda$CDM, if by other means is not possible. Nevertheless, it is true that the determination of $j$ imply measurements at larger redshifts than needed for $H$, which is an observational challenge harder than the determination of $H(z)$ \cite{sahni02, bamba18}.
The analysis performed for all the previous cases shows that the redshift evolution of the deceleration parameter $q$ still very close to the cosmological constant (the difference is much smaller than the accuracy of current observations). However assuming that $a_s=3$, then there is only one case, ($\alpha=3,\beta=-2$) corresponding to the EoS $p=-\rho^2/A$, where the current value of $q$ is $q_0\approx -0.6$ while $q_0^{\Lambda CDM}\approx -0.55$. In all cases the difference with $\Lambda$CDM increases towards the future. Considering the jerk parameter, the difference with $\Lambda$CDM is a little bit more marked. In Fig. 3 we show $j$ for the model ($\alpha=3,\beta=-2$), where the difference with $\Lambda$CDM is more appreciable, $j_0-j^{\Lambda CDM}\approx 0.25$.
\begin{center}
\includegraphics [scale=0.7]{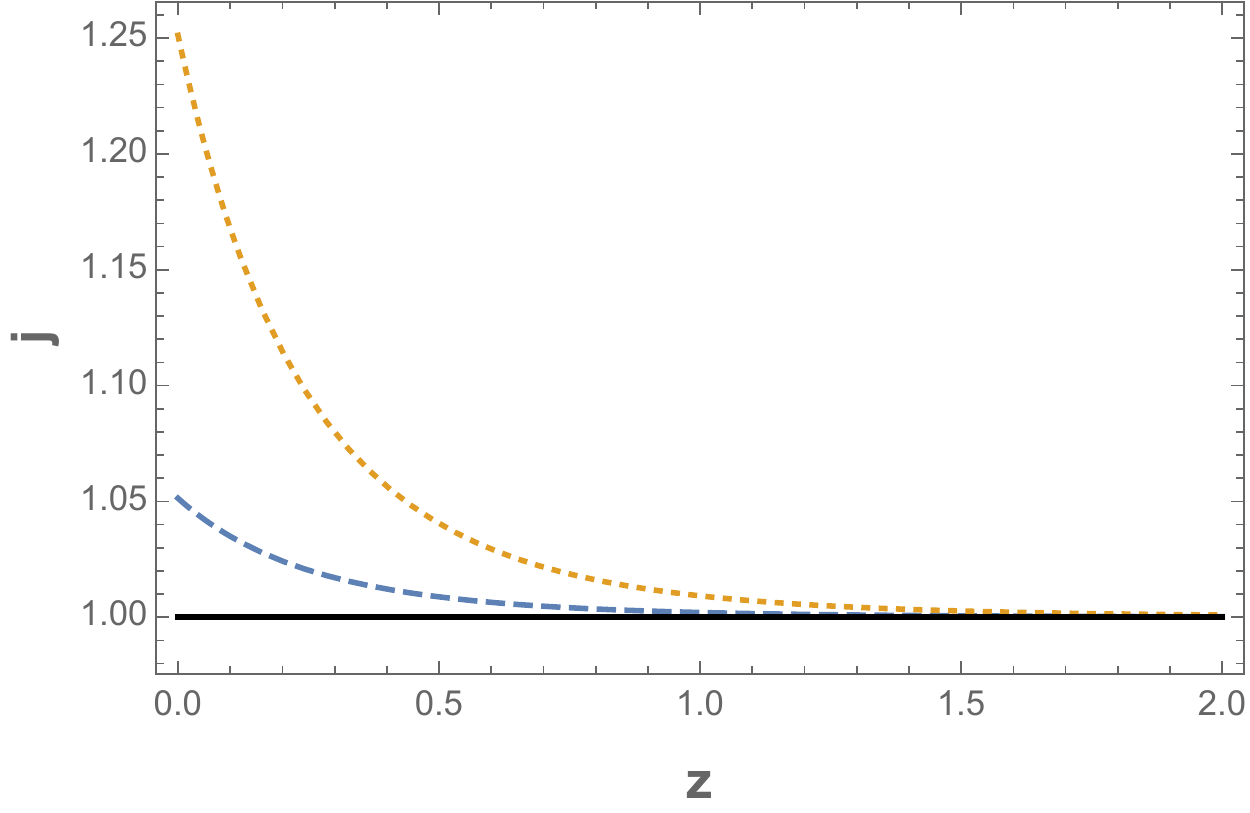}
\end{center}
\begin{center}
{Fig. 3 \it The jerk parameter for the model (\ref{eq13}), assuming $a_s=5$ (dashed) and $a_s=3$ (dotted). The constant $j=1$ corresponds to the $\Lambda$CDM model.} 
\end{center}
According to this behavior for $j$, the difference with $\Lambda$CDM increases at low redshift, being larger at the present. The current value for $a_s=3$ is $j_0\approx 1.25$, which falls between the range of current observational studies \cite{bamba18, zhong}. The time to the singularity for $a_s=3$ is $t-t_0\approx H_0^{-1}$. An important feature of the model (\ref{eq13}) is that the expressions for the scalar field and potential, for the DE component, are exact and valid for any $a/a_s<1$.
\subsection*{Type IV Singularities}
In type IV singularities the density becomes finite at $a_s$, which leads to non-singular EoS at $a_s$, but becomes singular at $a_s$ for $\ddot{H}$ and higher derivatives. Under appropriate choice of the parameters that respect the flatness condition and assuming an appropriate value for $a_s$, all these models lead to viable cosmological evolution and give a current value of the DE EoS in the range obtained from observations.
The following examples lead to type IV singularities, where the limits at $a\rightarrow a_s$ are given in Eq. (\ref{eq4k}), and will be shown in each case. We consider the following cases 

\noindent \textbf{1.} $\alpha=3$, $\beta=1/2$\\
\be\label{eq18}
p=-A\left(1-\frac{a^3}{a_s^3}\right)^{1/2},\;\;\; \rho=\frac{2}{3}A\sqrt{1-\frac{a^3}{a_s^3}}+\frac{2}{3}A\frac{a_s^3}{a^3}\left(1-\sqrt{1-\frac{a^3}{a_s^3}}\right)
\ee
In this case it is possible to find explicit dependence of the pressure in terms of the density as follows:
\be\label{eq18a}
p=-\frac{1}{4}\left(3\rho-2A+\sqrt{9\rho^2+12A \rho-12A^2}\right)
\ee
At $a\rightarrow a_s$ we find for $\rho$
\be\label{eq19}
\lim_{a \to a_s} {\rho}=\frac{2A}{3}.
\ee
The DE EoS is given by 
\be\label{eq19a1}
w_{DE}=-\frac{1}{4}\left(3-2\frac{A}{\rho}+\sqrt{9+12\frac{A}{\rho}-12\frac{A^2}{\rho^2}}\right)
\ee
\noindent  \textbf{2.} $\alpha=6$, $\beta=1/2$\\

\be\label{eq19c}
p=-A\left(1-\frac{a^6}{a_s^6}\right)^{1/2},\;\;\; \rho=\frac{A}{2}\left[\sqrt{1-\frac{a^6}{a_s^6}}+\frac{a_s^3}{a^3}\arcsin\left(\frac{a^3}{a_s^3}\right)\right]
\ee
The explicit dependence of the density in terms of the pressure is given by
\be\label{eq20}
\rho=-\frac{1}{2}p+\frac{A}{2}\frac{\arcsin\left(\sqrt{1-\left(\frac{p}{A}\right)^2}\right)}{\sqrt{1-\left(\frac{p}{A}\right)^2}}
\ee
The energy density tends to the limit at $a\rightarrow a_s$ 
\be\label{eq21}
\lim_{a \to a_s} {\rho}=\frac{A\pi}{4}
\ee
These models follow the same behavior as the previous models, except for the character of the singularity. In fact, the real singularity 
takes place for higher derivatives of $H$, ($|\ddot{H}|, |\dddot{H}|,...\rightarrow\infty$) at $t\rightarrow t_s$. It is clear that the jerk parameter  $j$ faces future singularity. The difference with the $\Lambda$CDM can be set only by this parameter. In Fig. 4 we show the jerk parameter for the model (\ref{eq18}) where the difference is more marked.
\begin{center}
\includegraphics [scale=0.7]{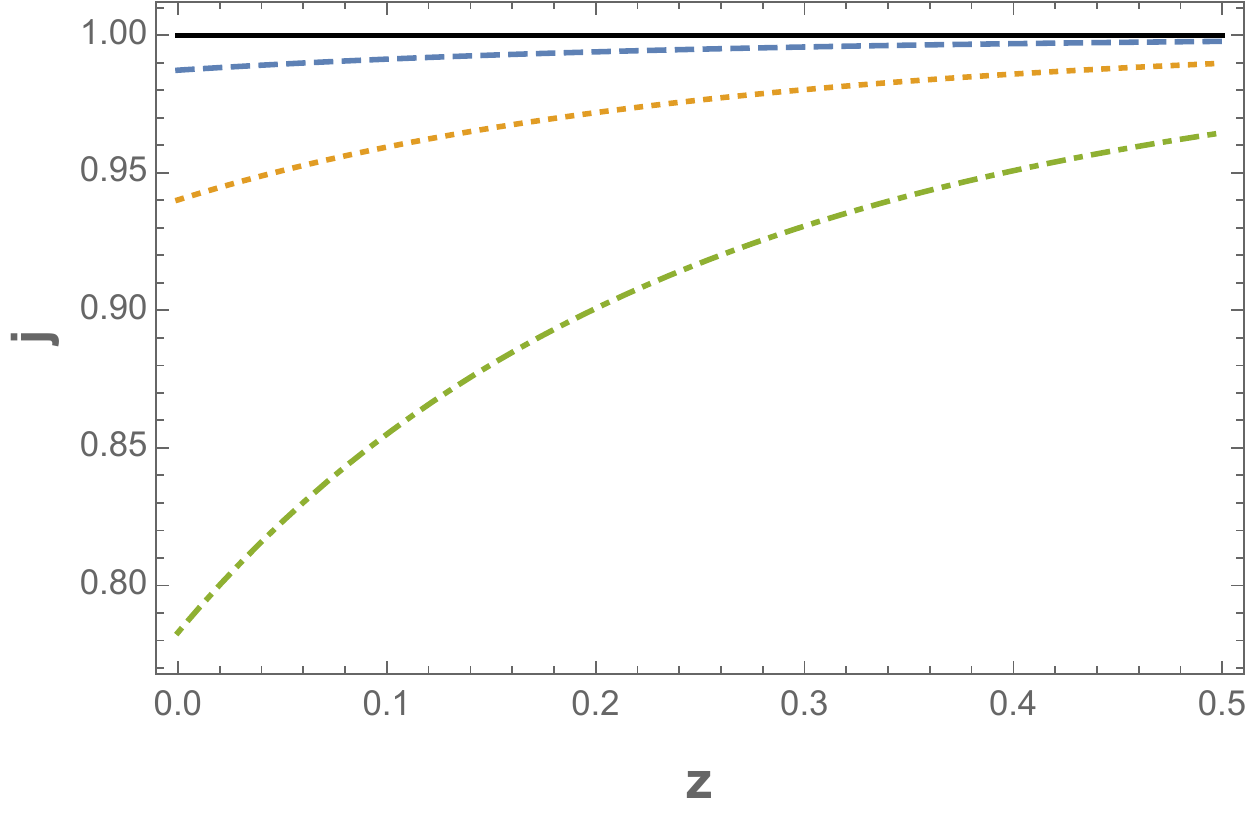}
\end{center}
\begin{center}
{Fig. 4 \it The jerk parameter for the model (\ref{eq18}), assuming $a_s=5$ (dashed), $a_s=3$ (dotted) and $a_s=2$ (dot-dashed). The remaining times to the singularity are $1.9 H_0^{-1}$, $1.3 H_0^{-1}$ and $0.8 H_0^{-1}$ respectively.} 
\end{center}

\section{Growth of Matter Perturbations}
With the improvement of the observational accuracy, more cosmological effects can be incorporated to the arsenal of tests that have the potential to discriminate between the jungle of dark energy models, allowing to constraint or rule them out. One of these effects has to do with the growth of matter perturbations in the universe, which depends on the expansion rate of the universe \cite{steinhardt}-\cite{huterer}. The dark energy acts against the growth of structure, and the greater the acceleration, the more suppressed the growth of structure, giving an important prove of DE effects with direct observations. For all the above considered models, the DE EoS varies very slowly with time, staying in an enough narrow interval around $-1$ through its evolution from matter dominance up to the present. This allows to apply the formalism of growth of matter density contrast $\delta=\delta\rho_m/\rho_m$ which, up to first order, satisfies the following equation \cite{linder2}
\be\label{eq40}
\ddot{\delta}+2H\dot{\delta}-4\pi G\rho_m\delta=0
\ee  
where $\rho_m$ is the background matter density and $\delta\rho_m$ is its first-order perturbation. The influence of the DE is encoded in the "friction term" proportional to $H$ in (\ref{eq40}). This equation was derived under the assumption that the scale of the perturbations is  smaller than the Hubble horizon, where the decoupling of dark matter and DE perturbations is a good approximation \cite{steinhardt,linder1, linder2}.
Defining the growth variable $g=\delta/a$ and using the $e$-folding variable $x=\ln a$, the equation (\ref{eq40}) takes the form \cite{linder1}-\cite{linder3}
\be\label{eq41}
\frac{d^2g}{dx^2}+\left[\frac{5}{2}-\frac{3}{2}w(a)(1-\Omega_m(a))\right]\frac{dg}{dx}+\frac{3}{2}\left(1-w(a)\right)\left(1-\Omega_m(a)\right)g=0
\ee
where $\Omega_m(a)=H_0^2\Omega_{m0} a^{-3}/H^2$ ($\Omega_{m0}$ is the current matter density parameter) and $w(a)$ is the DE EoS. A good approximation to the exact solution of this equation has the following analytical from
\be\label{eq42}
g(a)=\exp\left(\int_0^a \frac{da}{a}\left[ \Omega_m(a)^{\gamma}-1\right]\right)
\ee
where the parameter $\gamma$ is called the growth index, which with high accuracy can be considered as constant for $\Lambda$CDM and for a wide variety of DE models with slowly varying EoS \cite{linder2, linder3, dent}. Thus, for quintessence and phantom models the following expressions were found
\be\label{eq43}
\begin{aligned}
&\gamma=0.55+0.05\left[1+w(z=1)\right],\;\;\; w>-1\\ &
\gamma=0.55+0.02\left[1+w(z=1)\right],\;\;\; w<-1
\end{aligned}
\ee
So, even though the equation (\ref{eq41}) can be integrated numerically, the growth index $\gamma$ brings us closer to a way of parametrizing the effect of the DE in the evolution of structures in the universe.
All the models presented here give slowly varying EoS, even up to the near future before the (sudden or type III) singularity. In Figure 5 we show the growth function for the quintessence and Big Rip models derived from (\ref{eq1A3}).
\begin{center}
\includegraphics [scale=0.7]{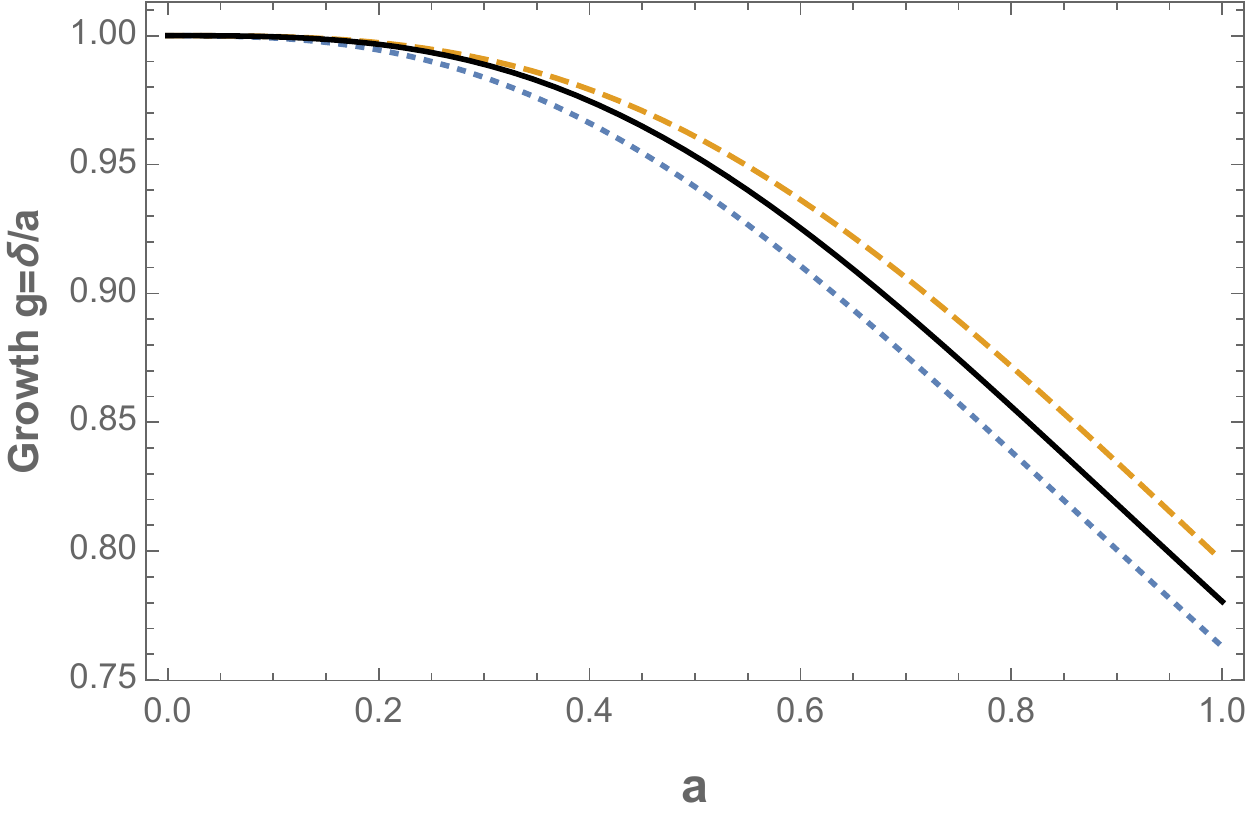}
\end{center}
\begin{center}
{Fig. 5 \it The growth function $g(a)$ for the quintessence and phantom model with Big Rip singularity derived from (\ref{eq1A3}). The (dotted) quintessence model corresponds to $\alpha=-1/2$ and $\beta=-1/2$, the phantom model (dashed) to $\alpha=3$ , $\beta=-1/2$ and the solid line corresponds to $\Lambda$CDM.} 
\end{center}
On large scales (larger than about 100 Mpc \cite{dent,dent1}) the above approximation loses rigor and scale-dependent correction is necessary, but this effect is more accentuated for modified gravity \cite{huterer, polarski}.
Thus, we assumed here that the growth index is quasi-constant for $\Lambda$CDM and so for models with smoothly varying EoS \cite{linder2, linder3}. Note that in all DE models and specially in models with future finite time singularity the matter component becomes increasingly subdominant towards the future, making the matter perturbations more and more imperceptible.
For the sudden and type III singularities the behavior of the growth function is very similar to the phantom case of Fig. 1. Another indicator of the effect of DE in the growth of density perturbations is the derivative of the logarithm of the density contrast $\delta$ with respect to logarithm of the cosmic scale 
\be\label{eq44}
f=\frac{d\ln \delta}{dx}
\ee
which, as follows from (\ref{eq40}) or (\ref{eq41}) satisfies the equation 
\be\label{eq45}
\frac{df}{dx}+\left(2+\frac{1}{2}\frac{d\ln H^2}{dx}\right) f-\frac{3}{2}\Omega_m(a)=0
\ee
which has an approximate solution of the form \cite{steinhardt, polarski}
\be\label{eq45}
f(a)=\Omega_m(a)^{\gamma}
\ee
with $\gamma\approx 6/11$ for the cosmological constant, this solution is also valid for slowly varying EoS in a narrow interval around $-1$. 
Applying this solution to the models of Fig.5 we find $f(a)$ as depicted in Fig.6
 \begin{center}
\includegraphics [scale=0.7]{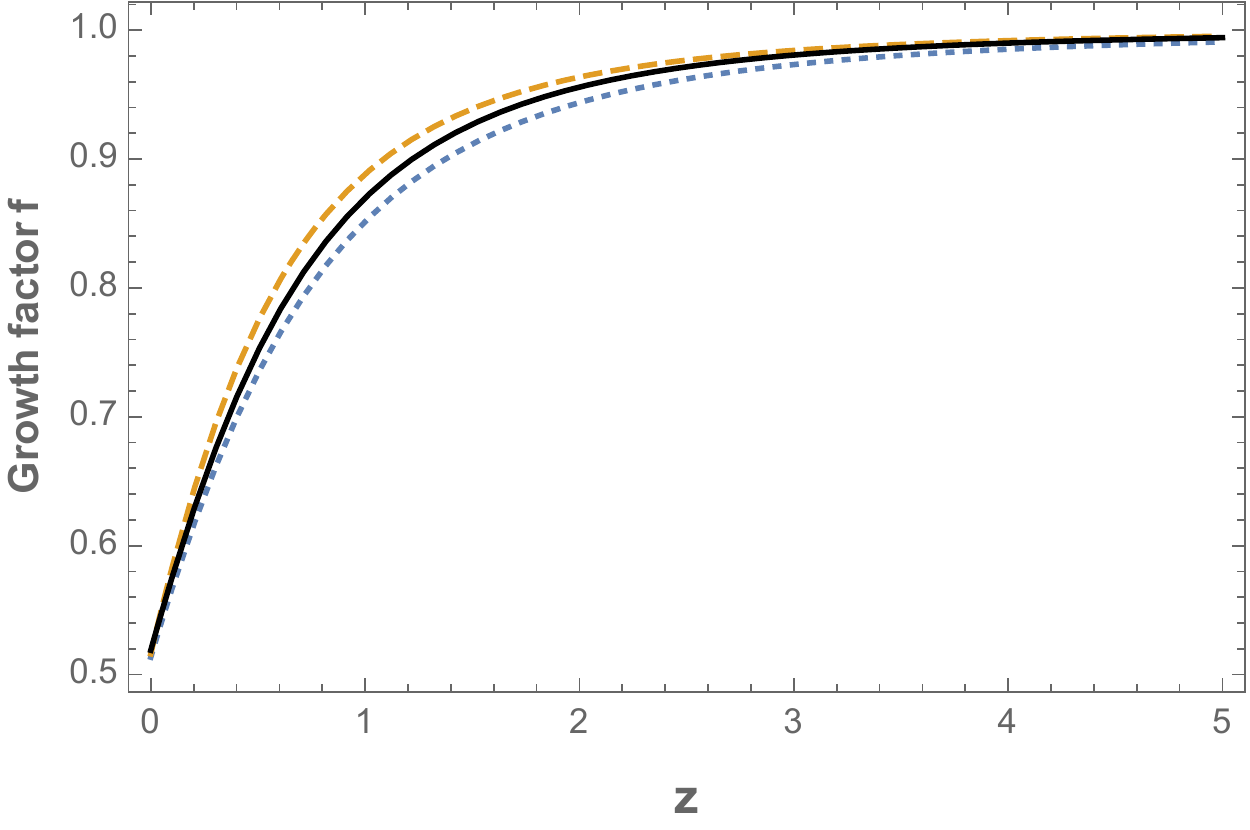}
\end{center}
\begin{center}
{Fig. 6 \it The growth factor $f(a)$ for the quintessence and phantom model with Big Rip singularity derived from (\ref{eq1A3}). The (dotted) quintessence model corresponds to $\alpha=-1/2$ and $\beta=-1/2$, the pantom model (dashed) to $\alpha=3$ , $\beta=-1/2$ and the solid line corresponds to $\Lambda$CDM.} 
\end{center}
The difference between the models is more noticeable in the redshift region $z\sim 0.5 -z\sim 1.5$, but it's still a subtle difference which needs more accurate observational data. So the future increment of observations in the redshift interval $(0.5,1)$ could play an important role in establishing a difference between $\Lambda$CDM and dynamical $\Lambda$CDM-like models, using the growth factor $f$.
\section{Discussion}
Assuming the probable fact that the dark energy is due to the negative pressure of an unknown type of matter, it seems appropriate to introduce models of DE by considering the pressure as function of the scale factor. This is specially interesting when studying cosmological scenarios with finite scale factor singularities and also to study exact solutions with Big Rip singularities and quintessence solutions. Besides that, this approach allows to directly follow the evolution of the main cosmological quantities in the observable redshift variable $z$. The model (\ref{eq1A3}), leads to two different families of DE models, depending on the parameters $\alpha$ and $\beta$. The first family is obtained for $\alpha>0$, $\beta<0$ and is equivalent to the scalar field model (\ref{eq1A4})-(\ref{eq1A5}), which describes phantom DE with  EoS given by (\ref{eq1A9}). In these models the universe evolves towards a Big Rip singularity at the time given by (\ref{eq1A7}). As the scale factor ($a\rightarrow\infty$), the EoS tends to $w\rightarrow -1-\alpha/3$. The second family of solutions, corresponding to the restrictions $\alpha<-3$, $\beta>0$ or $-3<\alpha<0$, $\beta<0$, are free of singularities and are reproduced by the quintessence scalar field model (\ref{eq1A10}) and (\ref{eq1A11}). An interesting feature of the potentials (\ref{eq1A5}) and (\ref{eq1A11}) is that retaining the first power in the scalar field give $V\sim const. +\phi^2$ which contains an exact Little Rip solution for a pressure of the form $p=-\rho-const.$ \cite{astashenok}.
The general model with pressure and density given by (\ref{eq4f}) and (\ref{eq4g}) gives different finite $a$ future singularities, depending on the chosen intervals for $\alpha$ and $\beta$.  
Thus, the sudden singularity takes place for $\alpha>0$ and $-1<\beta<0$. All the models in this case are indistinguishable from the cosmological constant up to the present, and even in the future before the singularity, and the farther the singularity the closer is the model to the cosmological constant. The restrictions $\alpha>0$, $\beta\le -1$ lead to type III singularities where two cases can be highlighted. The case $\alpha=3$, $\beta=-2$ gives the pressure $p=-\rho^2/A$ and allows the integration of the equations for the equivalent scalar field model with potential (\ref{eq14c}). 
The case $\alpha=1$, $\beta=-4$ gives the relationship $p=-\rho^{4/3}/A^{1/3}$ and is described by the scalar field (\ref{eqNA4}) with potential (\ref{eqNA5}). This model gives a simple connection between the current value of the EoS and the redshift at the singularity, $w_0=1/z_s$, which unambiguously allows to determine the time to the singularity by observational data on $w_0$.  As in previous cases of sudden singularity, the first three models of type III singularity are practically indistinguishable from the $\Lambda$CDM up to the present, but the last model ($\alpha=1$, $\beta=-4$) may differ appreciably from the cosmological constant at low redshift, which is interesting from the observational point of view. 
The same behavior, very similar to the $\Lambda$CDM, is observed for the examples of type IV singularities. This multiplicity of models that mimic the $\Lambda$CDM evolution can be reduced by using cosmological parameters that depend on higher order derivatives of the Hubble parameter, and can make a difference with the cosmological constant. The simplest of them is the jerk parameter which is constant for the $\Lambda$CDM model, $j=1$. In Fig. 3 we show the jerk parameter for the model ($\alpha=3, \beta=-2$)  where the difference with the $\Lambda$CDM is more accentuated, except for the model ($\alpha=1, \beta=-4$) which may differ from $\Lambda$CDM at the level of the EoS, as shown in Fig. 2. In Fig. 4 we plot the jerk parameter for the type IV singularity model ($\alpha=3, \beta=1/2$). Unfortunately with the available observational data is still difficult to bound the jerk parameter with sufficient reliability. It's worth noting that in all models we can bring the singularity closer to the future, and thus make a little difference with the $\Lambda$CDM. On the other hand, the growth of matter density perturbations also provides a useful tool to to test theoretical models of DE, thereby giving an insight on the fundamental physics. An interesting criteria to catch the effect of the DE on the growth of structure is given by the growth functions $g(a)$ and $f(a)$, which are characterized by the, phenomenologically obtained, growth index $\gamma$. Figs. 5 and 6 show the behavior of $g$ and $f$ for the models derived from (\ref{eq1A3}), compared to $\Lambda$CDM. It can be seen that for the function $g$, the difference with $\Lambda$CDM is more noticeable in the redshift interval $(z\sim 0,z\sim 1)$, while for $f$ the difference is more noticeable in the interval $(z\sim 0.5 ,z\sim 1.5)$. Thus, with the increase of observational data in the redshift interval $(z\sim 0 ,z\sim 1.5)$ these indicators will become more important.\\
The approach proposed in this paper uses the scale factor as the main variable to analyze the cosmological evolution of the universe, allowing to connect directly the measurable cosmological magnitudes with the observable redshift. Starting from the  expression (\ref{eq4f}) for the pressure and using the continuity equation, there were obtained, under appropriate  restrictions on the constants $\alpha$ and $\beta$, several physically different models that behave like the $\Lambda$CDM through all the evolution from the past to the present, but that make different predictions in the distant future (de Sitter, Big Rip, sudden, type III, type IV singularities). We first considered the expansion of (\ref{eq4f}) and (\ref{eq4g}) up to the power $(\frac{a}{a_s})^{\alpha}$, but it turned out that the resulting pressure and density satisfy the continuity equation, leading to an exact model that gives quintessence and phantom (with Big Rip) solutions. Then the finite scale factor singularities were obtained from (\ref{eq4f}) by the appropriate restrictions on $\alpha$ and $\beta$. This degeneracy, in the sense that physically different models are compatible with current observations, could be reduced as soon as the improvements in the observations allow to measure the cosmological parameters with higher accuracy.
\section*{Acknowledgments}
This work was supported by Universidad del Valle under project CI 71074 and by COLCIENCIAS grant number 110671250405.


\begin{thebibliography}{99}   
\bibitem{riess} A.G. Riess, et al., Astron. J. 116, 1009 (1998); astron. J. 117,
707 (1999). 
\bibitem{perlmutter} S.Perlmutter \textit{et al}, Nature \textbf{391}, 51 (1998)
\bibitem{kowalski} M. Kowalski, et al., Astrophys. Journal, \textbf{686}, p.749 (2008), arXiv:0804.4142
\bibitem{hicken} M. Hicken et al., Astrophys. J. 700, 1097 (2009);
[arXiv:0901.4804 [astro-ph.CO]].
\bibitem{komatsu} E. Komatsu et al. [WMAP Collaboration], Astrophys. J.
Suppl. \textbf{180}, 330 (2009); arXiv:0803.0547 [astro-ph].
\bibitem{percival} W. J. Percival et al., Mon. Not. Roy. Astron. Soc. \textbf{401}, 2148 (2010); arXiv:0907.1660 [astro-ph.CO]
\bibitem{komatsu1} E. Komatsu et al. [WMAP Collaboration], Astrophys. J. Suppl. \textbf{192}, 18 (2011) [arXiv:1001.4538 [astro-ph.CO]].
\bibitem{caldwell} R.R. Caldwell, Phys. Lett. B \textbf{545} (2002) 23; [arXiv:astro-ph/9908168]
\bibitem{sergei21} S. Nojiri, S. D. Odintsov and S. Tsujikawa, Phys. Rev. D \textbf{71}, 063004 (2005) [arXiv:hep-th/0501025].
\bibitem{bamba} K. Bamba, S. Capozziello, S. Nojiri, S. D. Odintsov, arXiv:1205.3421 [gr-qc]
\bibitem{frampton} P. H. Frampton and T. Takahashi, Phys. Lett. B\textbf{557}, 135 (2003) [arXiv:astro-ph/0211544].
\bibitem{kamion} R. R. Caldwell, M. Kamionkowski, and N. N. Weinberg, Phys. Rev. Lett. \textbf{91}, 071301 (2003) [arXiv:astro-ph/0302506].
\bibitem{barrow} J. D. Barrow, Class. Quant. Grav. 21, L79 (2004); [arXiv:gr-qc/0403084].
\bibitem{barrow1} J.D. Barrow, Class. Quant. Grav. 21, 5619 (2004); [arXiv:gr-qc/0409062]
\bibitem{sergei22} S. Nojiri and S. D. Odintsov, Phys. Rev. D\textbf{70}, 103522 (2004) [arXiv:hep-th/0408170].
\bibitem{stefancic} H. Stefancic, Phys.  Rev.D \textbf{71}, 084024 (2005); astro-ph/0411630
\bibitem{frampton1} P. H. Frampton, K. J. Ludwick, R. J. Scherrer, Phys. Rev. D\textbf{84}, 063003 (2011); arXiv:1106.4996v1 [astro-ph.CO]
\bibitem{brevik} I. Brevik, E. Elizalde, S. Nojiri, S.D. Odintsov, Phys.Rev. D\textbf{84} (2011) 103508; arXiv:1107.4642v2 [hep-th]
\bibitem{granda} L. N. Granda, E. Loaiza, IJMPD, \textbf{21}, (2012) 1250002; arXiv:1111.2454 [hep-th]
\bibitem{sergei24} S. Nojiri, S. D. Odintsov, Phys. Rev. D\textbf{70}, 103522 (2004); hep-th/0408170
\bibitem{sergei25} K. Bamba, S. D. Odintsov, L. Sebastiani, S. Zerbini, Eur. Phys. J. C\textbf{67}, (2010) 295; arXiv:0911.4390 [hep-th]
\bibitem{barrow2} J. D. Barrow, S. Cotsakis, A. Tsokaros,  Class. Quant. Grav. \textbf{27}, 165017  (2010); arXiv:1004.2681 [gr-qc]
\bibitem{ruth} L. P. Chimento, R. Lazkoz, Mod. Phys. Lett. A\textbf{19}, (2004) 2479; arXiv:gr-qc/0405020
\bibitem{astashenok} A. V. Astashenok, S. Nojiri, S. D. Odintsov, R. J. Scherrer, Phys. Lett. B \textbf{713}, (2012) 145; arXiv:1203.1976v1 [gr-qc]
\bibitem{nojiri} S. Nojiri, S. D. Odintsov, Phys.Rev. D\textbf{72}, 023003 (2005); arXiv:hep-th/0505215.
\bibitem{dutta} S. Dutta , R. J.Scherrer, Phys. Lett. B\textbf{676}, (2009) 12. 
\bibitem{cpl1} M. Chevallier, D. Polarski, Int. J. Mod. Phys. D, \textbf{10} (2001),  213; arXiv:gr-qc/0009008
\bibitem{cpl2} E.V. Linder, Phys. Rev. Lett., \textbf{90} (2003), 091301; arXiv:astro-ph/0208512
\bibitem{sahni02} V. Sahni, T. D. Saini, A. A. Starobinsky, U. Alam, JETP Lett. \textbf{77}, 201 (2003); arXiv:astro-ph/0201498
\bibitem{bamba18} A. Al Mamon, K. Bamba, arXiv:1805.02854 [gr-qc]
\bibitem{zhong} Zhong-Xu Zhai, et. al., Phys. Lett. B \textbf {727}, (2013) 8.
\bibitem{steinhardt} L. Wang, P. J. Steinhardt, Astrophys. J. \textbf{508}, 483 (1998); arXiv:astro-ph/9804015
\bibitem{linder1} E. V. Linder, A. Jenkins, 	Mon. Not. Roy. Astron. Soc., \textbf{346} (2003) 573; arXiv:astro-ph/0305286
\bibitem{linder2} E. V. Linder, Phys. Rev. D\textbf{72}, 043529 (2005); arXiv:astro-ph/0507263
\bibitem{linder3} E. V. Linder, R. N. Cahn, Astropart. Phys. \textbf{28}, 481 (2007); arXiv:astro-ph/0701317
\bibitem{dent} J. B. Dent, S. Dutta, L. Perivolaropoulos, Phys. Rev. D\textbf{80}, 023514 (2009); arXiv:0903.5296 [astro-ph.CO]
\bibitem{huterer} D. Huterer, et al., Astroparticle Physics \textbf{63}, 23 (2015); 
\bibitem{dent1} J. B. Dent, S. Dutta, Phys. Rev. D \textbf{79}, 063516 (2009); arXiv:0808.2689 [astro-ph]
\bibitem{polarski} R. Gannouji, D. Polarski, Phys. Rev. D \textbf{98}, 083533 (2018); arXiv:1805.08230 [astro-ph.CO]
\end{thebibliography}
\end{document}